\documentclass[aps,prl,reprint,10pt,longbibliography,superscriptaddress,nofootinbib]{revtex4-2}

\usepackage{amsmath, amssymb, amsfonts, amsthm, bm, mathrsfs, bbm, enumerate,mathtools}
\usepackage{graphicx}
\usepackage{epstopdf}
\usepackage{color}
\usepackage[usenames,dvipsnames]{xcolor}
\usepackage[citecolor=blue, colorlinks=true, urlcolor=blue, linkcolor=blue]{hyperref}
\usepackage{bbold}
\usepackage{amssymb}  
\usepackage{feynmp-auto} 
\usepackage{tikz}
\usepackage{soul}
\usepackage{float}
\usepackage{physics}

\usepackage{color}
\usepackage{nicefrac}
\usepackage{dsfont} 
\usepackage{wasysym} 
\usepackage{stmaryrd} 
\usepackage{hyperref} 
\usepackage{xcolor}
\usepackage{empheq} 

\definecolor{airforceblue}{rgb}{0.36, 0.54, 0.66}

\usepackage{verbatim}

\usepackage{cancel}
\usepackage{url}
\usepackage{pifont}
\newcommand{\cmark}{\ding{51}}  
\newcommand{\xmark}{\ding{55}}  

\newcommand{\mb}{\mathbf}

\newcommand\varpm{\mathbin{\vcenter{\hbox{%
  \oalign{\hfil$\scriptstyle+$\hfil\cr
          \noalign{\kern-.3ex}
          $\scriptscriptstyle({-})$\cr}%
}}}}

\begin{document}

\newcommand{\ourtitle}{Strong-to-weak symmetry breaking in monitored dipole conserving quantum circuits}
\title{\ourtitle}

\newcommand{\TUM}{\affiliation{Technical University of Munich, TUM School of Natural Sciences, Physics Department, 85748 Garching, Germany}}
\newcommand{\MCQST}{\affiliation{Munich Center for Quantum Science and Technology (MCQST), Schellingstr. 4, 80799 M{\"u}nchen, Germany}}
\newcommand{\Princeton}
{\affiliation{Department of Electrical and Computer Engineering, Princeton University, Princeton, New Jersey 08540, USA}}
\newcommand{\PQI}
{\affiliation{Princeton Quantum Initiative, Princeton University, Princeton, New Jersey 08540, USA}}

\author{Caterina Zerba}  \TUM \MCQST
\author{Sarang Gopalakrishnan}  \Princeton \PQI
\author{Michael Knap}  \TUM \MCQST

\begin{abstract}
We explore the information-theoretic phases of monitored quantum circuits subject to dynamics that conserves both charge and dipole moment, as well as measurements of the local charge density. Explicitly, both charge and dipole-moment conservation are strong symmetries, but under the dynamics they can be spontaneously broken to weak symmetries: this spontaneous symmetry breaking has an information-theoretic interpretation in terms of whether one can learn global charges from local measurements. We find a rich phase diagram: in one spatial dimension, charge is always easy to learn, while dipole moment can be either easy or hard. In two dimensions, we find three phases: for frequent measurements, both charge and dipole moment are easy to learn; as the measurement rate is decreased, first dipole moment and then charge become hard. In two dimensions, the low-measurement phase is an exotic critical phase with anisotropic spacetime scaling, analogous to a smectic liquid crystal.


\end{abstract}

\date{\today}

\maketitle

\textbf{Introduction.--}
Mixed states, subject to measurement and decoherence, can exhibit forms of order, as well as phase transitions, that are qualitatively different from pure states. The most famous example of a nontrivial mixed-state phase is a quantum error correcting code, which preserves a quantum memory so long as the error rate is below the error-correction threshold. However, in systems with symmetries, even simpler examples for mixed-state phases have been found~\cite{Buca_2012, Albert_2014, bao_symmetry_2021, Lee2023, Ogunnaike2023, Moudgalya_2024, sala_spontaneous_2024, Lessa2025, gu2024spontaneoussymmetrybreakingopen, Kuno2024, moharramipour_symmetry_2024, li_highly-entangled_2024,  Ma2025, Chen_2025, Huang_2025, lee_symmetry_2025, Weinstein_2025, Feng_2025, Ziereis2025}. In contrast to pure states, the notion of symmetries of mixed-state density matrices is more involved. A density matrix $\rho$ is said to be \emph{weakly symmetric} if its ensemble is invariant under the action of any element of the symmetry group on average, while the individual states are allowed to break the symmetry, $U\rho U^\dagger=\rho$. Intuitively, this can arise when the density matrix has contribution from different symmetry sectors. By contrast, \emph{strong symmetry} requires each state in the ensemble to be symmetric, leading to the more stringent condition $U\rho=e^{i\theta}\rho$ for some $\theta$. This can arise when all of the states in the ensemble originate from the same symmetry sector. 

\begin{figure}
    \centering
    \includegraphics[width=0.98\linewidth]{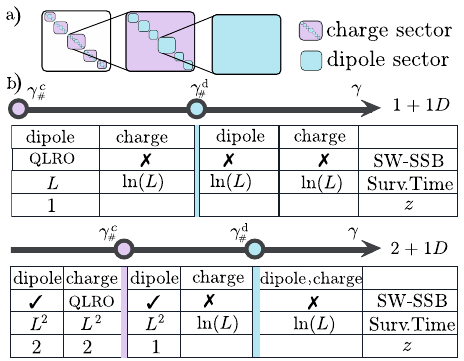} \caption{\textbf{Non-equilibrium phases in monitored dipole-moment conserving circuits.} a) Illustration of hierarchical strong-to-weak spontaneous symmetry breaking (SW-SSB): (left) both charge and dipole sectors are mixed and hence weakly symmetric; (middle) charge is strongly symmetric while dipole is weakly symmetric; (right) both charge and dipole are strongly symmetric. b) The hierarchy of non-equilibrium phases in monitored quantum circuits realized by tuning the measurement rate $\gamma$ (top: 1+1D, bottom: 2+1D). We characterize the non-equilibrium phases by SW-SSB, survival time (i.e., sharpening time) of a weakly-symmetric initial state in a finite system of linear extent $L$, and the dynamical exponent $z$.
    Here, \cmark{} indicates the weakly symmetric (SW-SSB) phase, \xmark{} the strongly symmetric phase, and quasi long-range order (QLRO) power-law decaying correlations. 
    For the gapped degrees of freedom, $z$ is left blank.
    }
    \label{fig:1}
\end{figure}
Monitored circuits realize transitions between different non-equilibrium phases as a function of the measurement rate~\cite{ Skinner2019,Li2018,Szyniszewski2019,Li2019,Nahum2020,Zabalo2020,Bao2020, Jian2020, Gullans2020aug,Gullans2020oct,Choi2020,Chen2020, Fisher2023}. It has been shown, that a mixed state evolved through noisy monitoring channels with strong conservation of U(1) charges can for small measurement rates exhibit strong-to-weak spontaneous symmetry breaking (SW-SSB) up to channel depths that scale polynomially in system size~\cite{singh2025, Vijay2025, Hauser2025}. By contrast, a high measurement rate selects a single symmetry sector of an initially weakly symmetric ensemble in circuit depth of order $\log L$. Crucially, this transition is absent in the circuit- and measurement-averaged density matrix~\cite{Ziereis2025}. 
The transition can be equivalently interpreted as a transition in the learnability of quantum states on how fast local measurements extract information of conserved quantities~\cite{Agrawal2022,Barrat2022, Barratt2022, Majidy2023, Agrawal2024,Ha2024,Fava2024,Ippoliti2024,Feng2025chargespin,singh2025}.
Understanding the fate of these non-equilibrium phases, especially in systems with unconventional conservation laws beyond charge conservation, is largely an open question.

In this work, we study the non-equilibrium phases of monitored quantum circuits that strongly conserve not only $U(1)$ charge, but also its associated dipole moment. These systems exhibit fractonic behavior and subdiffusive transport~\cite{Feldermaier2020, Gromov2020, Sanchez2020, Sala19, khemani20192d}, and can host different hierarchical symmetries of the density matrix; see Fig.~\ref{fig:1}(a).
Using field-theoretic techniques, we analyze SW-SSB in monitored dipole-conserving circuits in both 1+1 and 2+1 dimensions (D). We find that the system exhibits up to three distinct non-equilibrium phases as the measurement rate is increased, separated by either continuous or first-order transitions. In particular, the weak-measurement phase we find in two dimensions is an information-theoretic analog of the smectic phase in \emph{three-dimensional} liquid crystals, i.e., it is a nontrivial critical phase at its lower critical dimension~\cite{PhysRevLett.47.856, PhysRevLett.75.4752, radzihovsky2024critical}. 

Our results highlight some subtle distinctions in these information-theoretic problems that had not been previously appreciated. First, we find a phase (at intermediate measurement rates in 2D) for which the global charge is easy to learn, yet the local charge fluctuations conditional on the measurement outcome are still long-range, i.e., algebraically decaying. Naively, one might have expected easy learning to correspond to measurements ``fixing'' the charge profile of the conditional state. Second, we observe that the dynamical critical exponent $z$ (determined from correlation functions) need not determine the timescale $t(L)$ on which one learns the charge of a system of size $L$: we find a pair of phases with distinct values of $z$ but parametrically the same scaling of $t(L)$. Our results clarify the distinctions between these concepts, and illustrate the rich variety of phenomena that occur in monitored symmetric quantum matter.


\textbf{Model and observables.--}We consider evolution under random quantum circuits consisting of channels that preserve both the total charge and total dipole moment as strong conservation laws. These conserved charges are respectively 
$Q = \sum_{\boldsymbol{r}_i} n_{\boldsymbol{r}_i}$ and
$P = \sum_{\boldsymbol{r}_i} \boldsymbol{r}_i n_{\boldsymbol{r}_i}$, where $\boldsymbol{r}_i$ is the position of site $i$ in the circuit.
Between each layer of channels, we weakly measure the local charge $n_{\boldsymbol{r}_i}$ at some rate $\gamma$ at every site; equivalently, one could measure it projectively on some fraction $\gamma$ of sites. 
In either case, these measurement outcomes are recorded, giving a spacetime vector $\mathbf{m}$ that represents the measurement record on a particular run of the experiment. 

We will focus on the properties of the conditional state $\rho_{\mathbf{m}}$---e.g., whether its total charge and/or dipole moment are fixed given $\mathbf{m}$, and how its conditional spacetime correlation functions decay. 
We will discuss three classes of observables. First, following Refs.~\cite{Agrawal2022, Barrat2022, Barratt2022}, one can initialize the system in a state of uncertain charge, and compute the charge variance in the conditional state as a function of time: ``charge learning'' (or ``sharpening'') occurs when the measurement record $\mathbf{m}$ completely fixes the charge of $\rho_{\mathbf{m}}$. Second, we will consider properties of the steady state that results from taking an arbitrary initial state and running it for very long times. 
First, we will consider the order parameter for SW-SSB, by computing the R\'enyi-2 correlator for operators $O_x$ charged under the considered symmetry: 
\begin{equation} \label{eq:renyi2}
    \mathcal C_2(x-y, t) = \mathbb{E}_\mathbf{m} \mathbb{E}_u\bigg[\frac{\text{Tr}( \rho_{\mathbf m }(t)O_x^\dagger O_y\rho_{\mathbf m }(t)O_y^\dagger O_x)}{\text{Tr}(\rho_{\mathbf m }(t)^2)}\bigg],
\end{equation}
where $\mathbb{E}_{\mathbf{m}}(\cdot) \equiv \sum\nolimits_{\mathbf{m}} p_\mathbf{m} (\cdot)$ and $\mathbb{E}_u(\cdot)$ is the circuit average. 
We will consider this correlator in the steady state, i.e., taking $t\to \infty$ at finite size, then $L \to \infty$. In this limit, for any record $\mathbf{m}$, $\rho_{\mathbf{m}}(t)$ is strongly symmetric. The behavior of $\mathcal{C}_2(x-y)$ as $|x-y|\to \infty$ determines whether SW-SSB exists as long-range order or quasi-long range order, or is absent. 
In addition to this R\'enyi-2 correlator, we also compute other features of the conditional steady state, such as its connected density correlations $\langle n_{\mathbf{x}} n_{\mathbf{y}} \rangle_c \equiv \mathbb{E}_{\mathbf{m}}[ \mathrm{Tr}(n_\mathbf{x} n_\mathbf{y} \rho_\mathbf{m}) - \mathrm{Tr}(n_\mathbf{x} \rho_\mathbf{m}) \cdot \mathrm{Tr}(n_\mathbf{y} \rho_\mathbf{m})]$. 
These correlation functions can straightforwardly be generalized to unequal times~\cite{SM}.
A distinct quantity we will also consider is the dynamics of charge sharpening, starting from a state that explicitly breaks the strong symmetry (i.e., has an uncertain charge). The sharpening (or survival) time $t_{\#}(L)$ for a system of size $L$ is the timescale on which the dynamics restores the strong symmetry on individual trajectories. Note that charge sharpening can be interpreted as a temporal version of the R\'enyi-2 order parameter for SW-SSB: in the sharpening setup, one breaks the strong symmetry at the lower temporal boundary, and checks to see if it has been restored at the upper temporal boundary.

\begin{figure}
    \centering
    \includegraphics[width=0.98\linewidth]{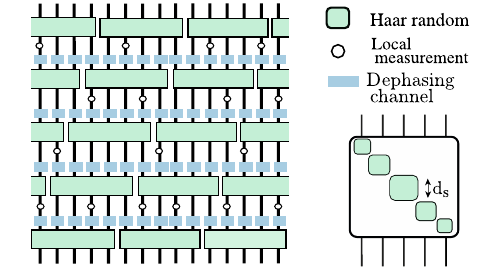}
    \caption{\textbf{Strongly symmetric quantum channel.} We consider a brickwork circuit composed of gates conserving the global charge and dipole moment. Each gate connects configurations within the same symmetry sector via Haar-random unitaries. The unitary evolution is followed by a completely dephasing channel, and local measurements are performed at each site with a certain rate $\gamma$. }
    \label{fig:2}
\end{figure}

It remains to specify the ensemble of channels. We choose these as follows: the ``core'' of each channel is a five-site unitary gate, arranged in a period-5 brickwork pattern (Fig.~\ref{fig:2}). Each such gate is bracketed by single-site dephasing channels: one could interpret these as measurements for which the outcome is not recorded. To avoid the emergence of dynamically disconnected Krylov subspaces within a fixed charge-dipole sector, resulting from Hilbert space fragmentation~\cite{Sala19, khemani20192d, Rakovszky20, Moudgalya_2022}, we choose sufficiently high interaction ranges of five sites of the dipole-conserving gates. A gate acting on $n$ neighboring sites starting at site $i$ takes the form $U_{\mathbf{r}} = \bigoplus_{s \in \mathcal{S}} U^{(s)} P_s$, where $\mathcal{S}$ is the set of connected charge-dipole sectors, $P_s$ is the projector onto sector $s$, and $U^{(s)}$ is a Haar-random unitary of dimension $d_s$. Here, $d_s$ is the size of the charge and dipole sector $s$ and counts the number of connected configurations. This construction naturally generalizes to systems with higher-moment (multipole) symmetries~\cite{Feldermaier2020} or other sets of constrained configurations.

\textbf{Effective evolution.--}We analyze an effective statistical mechanics description for the charge configurations averaged over the set of unitaries~\cite{Barratt2022, singh2025}. In the absence of measurements the evolution provided by the quantum channel composed of the averaged unitaries and dephasing channel can be described by a transfer matrix and is obtained by a standard calculation~\cite{SM}:
$\mathbb T_u = \mathbb{1}- \sum_{\alpha, \mathbf{r}} P^\alpha_{\mathbf{r}}  + \sum_{\alpha\mathbf{r}} \left\{ J_\alpha\, \mathrm{d}t \left(G^\alpha_{\mathbf r} + G^{\alpha\dagger}_{\mathbf r} \right) + (1 - J_\alpha \mathrm{d}t) \mathbb{1} \right\} P^\alpha_{\mathbf r}$, where $\alpha$ labels the distinct classes of gates, $J_\alpha$ are coupling constants proportional to the probability of applying the gates $G^\alpha_\mathbf{r}$ and $G^{\alpha\dagger}_\mathbf{r}$ acting on $\mathbf r$, and $P^\alpha_\mathbf{r}$ is the projector onto the subspace of these configurations. Exponentiating the expression, we obtain $\mathbb T_u \sim \exp\left( \mathrm{d}t \sum_{\alpha,\mathbf{r}}J_\alpha \left[ (G^\alpha_\mathbf{r} + G^{\alpha\dagger}_\mathbf{r}) P^\alpha_\mathbf{r} - P^\alpha_\mathbf{r} \right] \right) \equiv \exp(-H_u \mathrm{d}t )$, where we introduced the effective Hamiltonian $H_u$ governing the imaginary time evolution. Here the dephasing channel ensures that the transfer matrix is diagonal in the replica indices.
To compute the average of observables that are non-linear in the density matrix, we introduce different replicas, where in each of them the charge evolves according to the transfer matrix $T_u$ and is pinned by measurements, that forces the configuration to be the same for all replicas when monitored.
For the field-theoretic treatment we replace the discrete projective measurements by a smooth Gaussian operator that enforces measurement outcomes across all replicas with a soft constraint:
$
\mathbb T_m = \exp\left( -\frac{\gamma}{2}\, \mathrm{d}t \sum_{\mathbf{r}, a} \left( \sigma^z_{\mathbf{r}a} - m_{\mathbf{r}}(\tau) \right)^2 \right),
$
where $m_{\mathbf{r}}(\tau)$ is a continuous-valued measurement outcome in the interval $[-1,1]$~\cite{Barrat2022}. Integrating out the measurements, one obtains an effective total Hamiltonian description for the evolution of the charge in imaginary time,
$
H = H_u + H_m,
$
where 
$H_m = \sum_{\mathbf{r}, a, b} \sigma^z_{\mathbf{r}, a}\, \mathbb{P}_{a b}\, \sigma^z_{\mathbf{r}, b}$, with a projector $\mathbb{P}_{a b} = \delta_{a b} - \frac{1}{\mathcal{N}}$  onto the asymmetric component of the replica degrees of freedom. This effective statistical mechanics description on replicas is now amenable to field theoretic approaches.

\textbf{Field theory.--}On the technical level, we use a spin path integral formalism to derive an effective field theory for our model. We define the spin coherent state $\ket{\mathbf{n}} = e^{i \theta (\mathbf{n}_0 \times \mathbf{n}) \cdot \mathbf{S}}\ket{S,S}$, where the spin operators satisfy $\bra{\mathbf{n}} \mathbf{S} \ket{\mathbf{n}} = S \mathbf{n}$ and $\ket{S,S} $ is the eigenvector of $S^2, S^z$. The unit vector is parameterized as $\mathbf{n} = (\sin\theta\cos\phi, \sin\theta\sin\phi, \cos\theta)$, with $\mathbf{n}_0 = (0, 0, 1)$~\cite{Fradkin_2013}.
For a dipole-moment conserving model in generic $D$ dimensions, we expand around the largest charge sector $\theta=\pi/2+\vartheta$, corresponding to half filling, and construct an effective Lagrangian that includes all symmetry-allowed terms~\cite{Lake2022}:
  
\begin{align}
    \mathcal{L} &= \sum_a \Big[\frac{i}{2} \vartheta_a(\mathbf{r}) \partial_\tau \phi_a(\mathbf{r}) 
    + \frac{J}{4} \Big\{ \sum_i \left( \partial_i^2 \phi_a(\mathbf{r}) \right)^2 
    + \left( \partial_i^2 \vartheta_a(\mathbf{r}) \right)^2 \Big\} \nonumber \\
    & + \frac{J'}{4} \Big\{ \Big( \sum_i \partial_i^2 \phi_a(\mathbf{r}) \Big)^2 
    + \Big( \sum_i \partial_i^2 \vartheta_a(\mathbf{r}) \Big)^2 \Big\} \Big] \nonumber \\
    & + \frac{\gamma}{2} \sum_{a,b} \vartheta_a(\mathbf{r}) \mathbb{P}_{a,b} \vartheta_b(\mathbf{r}).
\end{align}
Here $J$ and $J'$ are coefficients that depend on the details of the connected configurations.  The fields $\phi, \vartheta$ can be decomposed in the two projectors over the replicas, as $\phi = \mathbb{P} \phi + \mathbb{Q} \phi$, where $\mathbb{P}$ is the asymmetric component introduced above, and  $\mathbb{Q} = \mathbb{1} - \mathbb{P}$ the symmetric one. Intuitively, we expect that  phase transitions in the measurement rates are encoded in the $\mathbb P$ component since measurements are effectively coupling replicas and that $\mathbb{Q}$ does not undergo a transition.

After integrating out the measurement terms $\mathbb P \vartheta $ and applying a long-wavelength approximation, we obtain that
the field $\mathbb P \phi $ is governed by a Lifshitz model, common in effective descriptions of dipole-conserving systems~\cite{Zhai2021, Lake2022, Moudgalya_2021, Zechmann2023, Lake2023, Boesl_2024, Zechmann_2024}. To gain further insight into the symmetry properties of mixed state for dipole and charge, we introduce an additional degree of freedom $\phi_d$ to represent the dipole phase, coupled to the charge phase $\phi$:
\begin{align}\label{lag2d}
    \mathcal{L} &=  \mathbb{Q} \Big[\frac{i}{2} \vartheta \partial_\tau \phi 
    + \frac{J}{4} \sum_i (\partial_i^2 \vartheta)^2 
    + \frac{J'}{4} \Big( \sum_i \partial_i^2 \vartheta \Big)^2 \Big] \nonumber \\
    & + \left( \frac{a_p}{2} \mathbb{P} + \frac{a_q}{2} \mathbb{Q} \right)
    [ K_{ijkl} \partial_i \phi_d^j\partial_j \phi_d^k] 
     + \left( \frac{b_p}{2} \mathbb{P} + \frac{b_q}{2} \mathbb{Q} \right) 
    (\partial_\tau \phi_d)^2 \nonumber \\
    & +  \frac{c_p}{2} \mathbb{P} 
    (\partial_\tau \phi)^2+ r \left( \frac{d_p}{2} \mathbb{P} + \frac{d_q}{2} \mathbb{Q} \right) 
    (\nabla \phi + \phi_d)^2.
\end{align}
The coefficients of the terms are circuit-specific parameters, explicitly obtained for the 1+1D case in the supplemental material \cite{SM}, and $r$ is a large auxiliary parameter that enforces the constraint $\phi_d = -\nabla \phi$ at low energies (long times). This ensures that dipole and charge fields remain formally independent in the theory, while the field theory gaps out all the configuration where the relation is not satisfied~\cite{Zhai2021,Lake2022,Moudgalya_2021, Zechmann2023, Lake2023, Boesl_2024, Zechmann_2024}. One can recover the  Lagrangian for the charge phase by integrating out $\phi_d$ in the $r \to \infty$ limit, or vice versa obtain an effective Lagrangian for $\phi_d$ by integrating out $\phi$. 

We now analyze the effective Lagrangian~\eqref{lag2d} and show that measurements can drive a transitions from a weakly to a strongly symmetric non-equilibrium phase. We separately study charge and dipole sectors and find that the resulting phase diagram varies qualitatively with the dimensionality of the system.


\textbf{Non-equilibrium phases in 1+1D.--}In 1+1D the Lagrangian~\eqref{lag2d} can be simplified to
   
\begin{align}\label{1Dmod}
    \mathcal{L}=&\mathbb Q \{\frac{i}{2} \vartheta\partial_\tau \phi + \frac{\bar \rho }{2} (\partial_x^2 \vartheta)^2 \}\mathbb Q+ \big[\big(\frac{\rho_s}{2} \mathbb P+ \frac{\bar \rho}{2}\mathbb Q \big)\partial_x \phi_d]\partial_x \phi_d +\nonumber \\& +  \big[\big(\frac{\rho_s}{2} \mathbb P+ \frac{\bar \rho}{2}\mathbb Q \big)\partial_\tau \phi_d]\partial_\tau \phi_d + \nonumber \\&+ r\big[\big(\frac{\rho_s}{2} \mathbb P+\frac{\bar \rho}{2}\mathbb Q \big)(\partial_x \phi +\phi_d)\big](\partial_x \phi +\phi_d),
\end{align}
where we have rescaled space and time to obtain an isotropic model with $\rho_s \propto 1/\gamma^{2/3}$ and $\bar \rho\sim J$~\cite{SM}. 
In order to determine the non-equilibrium phases, we analyze the vortex proliferation for the phase of the charge and the dipole, mapping the model to a modified sine-Gordon in the dual variable $\varphi$:  
\begin{align*}
\mathcal L&=\frac{1}{8 \pi^2 \bar\rho} \bigg[(\partial_x^4 \mathbb Q\varphi)^2+(\partial_\tau \mathbb Q\varphi)^2\bigg]+\frac{1}{8 \pi^2 \rho_s}\bigg[(\partial_x^2 \mathbb P\varphi)^2+\\& +(\partial_\tau \mathbb P\varphi)^2\bigg]-\sum_{a\not= b \,\, m} [ g_b^{(m)} \cos(m(\partial_x \varphi_a-\partial_x \varphi_b))+\\&+g_s^{(m)} \cos(m( \varphi_a- \varphi_b))],
\end{align*}
where $e^{i\varphi_a}$ ($e^{i\partial _x \varphi_a}$) inserts a vortex of the dipole (charge). The minimal vortex configurations belong to the replica asymmetric sector $\mathbb P$. By an entropic argument, we find that excitations that have a non-zero component in the symmetric sector $\mathbb Q$ are always bound and cannot drive a phase transition, as anticipated before.

In the asymmetric sector $\mathbb P$, the cosine operator associated with dipole vortices is irrelevant at the Gaussian fixed point  $g_b=g_s=0$. By contrast, the charge cosine operator  
is always relevant, which implies that the charge vortices are unbound for any measurement rate, and the RG flows to large values of $g_b$. As for large values of $g_b$ the field $\partial_x \varphi_a-\partial_x \varphi_b$ is locked in one of the minima of the cosine, we can expand the cosine term and obtain the effective sine-Gordon theory for the dipole vortices\begin{align*}
  \mathcal L &=  \sum_a \bigg[\frac{1}{8 \pi^2 \rho_s}(\partial_\tau \mathbb P\varphi_a)^2+\sum_{b\not= a} \frac{ g_b }{2} (\partial_x \varphi_a (\mb r)-\partial_x \varphi_b (\mb r))^2\\&-g_s \cos( \varphi_a (\mb r)-\varphi_b (\mb r)),
\end{align*} which we restricted to the $\mathbb P$ component. This model undergoes a Berezinskii-Kosterlitz-Thouless (BKT) transition from gapless to gapped dipoles, while the charge remains gapped throughout~\cite{Zechmann2023,Lake2023}. 
Intuitively, the proliferation of inter-replica dipole vortices causes large fluctuations of the inter-replica dipole phase. In this phase, the conjugate variable---inter-replica dipole \emph{density}, i.e., the difference in dipole density between replicas---is pinned to zero. Thus, the phase in which dipole vortices proliferate is the phase in which all replicas agree on local dipole density, i.e., the sharp phase. 

We can summarize the properties of the two phases as follows; see Fig.~\ref{fig:1}(b). For an in-depth analysis, we refer to the Appendix A and B. In the dipole-sharp phase, the R\'enyi-2 correlator, Eq.~\eqref{eq:renyi2}, decays exponentially, and so do correlation functions of dipole or charge density in the conditional state. The associated correlation length diverges exponentially at the dipole sharpening transition, which is a BKT transition. Moreover, starting from states that are not dipole-sharp, sharpening takes place on a timescale $\sim \log L$ for a system of length $L$. In the dipole-fuzzy phase, on the other hand, the R\'enyi-2 correlator decays algebraically with a correlation length that varies throughout the phase, indicating quasi-long-range order. Meanwhile, the correlator of the dipole \emph{density} $n_d=\partial_x \varphi$ decays algebraically, as $1/x^2$, and sharpening takes place on a timescale $\sim L$ (set by the gap of the single-trajectory transfer matrix). 

So far, the phases we have considered are precisely analogous to the standard $U(1)$ charge-sharpening problem. We now consider the properties of \emph{charge}. Naively, since charge is gapped throughout the phase diagram, one might expect all correlation functions associated with it to decay exponentially. This is indeed the case for the R\'enyi-2 correlator of the charge. Consequently, also, charge sharpens on a timescale $\sim \log L$ for any nonzero measurement rate. Notably, however, charge fluctuations do not decay exponentially in the conditional state: since the charge can be written as $n_c=\partial^2_x \varphi$, in terms of the dipole density, its correlation functions decay as $1/x^4$ in the dipole-fuzzy phase. Despite being algebraic, these fluctuations are sufficiently suppressed that the charge variance in a region of size $\ell$ does not scale with $\ell$---thus, they are consistent with the fact that the charge profile is rapidly learned.


\textbf{Non-equilibrium phases in 2+1D.--}From the analysis of the 1+1D model we saw that the excitations that drive the phase transitions are the ones that have zero overlap with the $\mathbb Q$ space.  
As such, we can understand the phase diagram for 2+1D by looking only at the spatial decay of vortex excitations in the $\mathbb P$ components of the fields, that are described by a Lifshitz model of Eq.~\eqref{lag2d}. 
The equilibrium phases described by the Lifshitz model have for example been analyzed in Refs.~\cite{Radzihovsky2022, Lake2022}.

Generalizing these considerations to our problem, we find that the model in Eq.~\eqref{lag2d} predicts three possible phases; see Fig.~\ref{fig:1}(b), and Appendix A, B. When measurements are sufficiently weak or sparse, both dipole and charge obey SW-SSB, with long-range order for the dipole moment and quasi-long-range order for the charge. In this phase, the sharpening time scales as $t^{c,d}_{\#} \sim L^2$. This scaling for the charge sharpening time is due to the presence of critical charge fluctuations with dynamical scaling exponent $z = 2$. The dipole moment also sharpens on this timescale: this timescale is related to the length-scale for loss of order in a three-dimensional smectic that is finite along its two transverse directions and infinite along the third~\cite{PhysRevLett.75.4752}. Because this phase is governed by a critical theory with $z = 2$, the spatio-temporal density correlation functions in the conditional state exhibit scaling with the same exponent. 



Increasing the measurement rate, charges become gapped. The phase transition is expected to be analogous to a smectic to a nematic-A phase transition in 2+1D, once the imaginary time is identified with the direction perpendicular to the plane~\cite{Lake2022}. 
In this regime, the phase of the dipole is still condensed. 
Dipoles exhibit SW-SSB with long-range order; this phase is a conventional Goldstone phase in which the Goldstone modes have a linear dispersion, so $z = 1$. Accordingly, spacetime correlations of the dipole density in the conditional state exhibit dynamical scaling with $z = 1$. Nevertheless, the sharpening time scales as $t_{\#, d} \sim L^2$. 
The sharpening time is obtained from determining the minimal gap of the transfer matrix, which here is not given by the collective modes but by Anderson's tower of states instead~\cite{Anderson1952}. At the same time $t^c_{\#} \sim \log L$, as charge sectors are gapped.

Increasing the measurement rate even further, leads to a third regime, in which both the dipole sector and the charge sector are efficiently selected by the monitored evolution on timescales $t^{c,d}_{\#}\sim \log L$, resulting in a strongly symmetric charge and dipole phase. This transition is generally understood to be second-order, associated with dipole condensation. However, depending on the details of the circuit, the transition could be rendered first order as well. In particular, this happens when interactions between dipoles are unbound from below to quartic order, as seen in Ref. \cite{Lake2022}. 

In the supplement, we further discuss the density fluctuations of charge and dipole, that similarly to 1+1D are not given the gap of respective sectors of the transfer matrix, as the charge and dipole density are related by a derivative~\cite{SM}.

\textbf{Discussion and outlook.--}In this work, we studied a monitored dipole-conserving model which realizes different non-equilibrium phases. These phases are diagnosed by strong and weak symmetries of charge and dipole, respectively. The phases are also identified by the depth of the circuit needed to determine the underlying charge structure. Phase transitions between these phases are identified as charge sharpening transitions. 
Our results highlight the interplay between measurements and unconventional symmetries, demonstrating that dipole-conserving quantum channels can exhibit SW-SSB in averaged correlation functions that are non-linear in the density matrix. Furthermore, we highlight that the gap of the respective transfer matrix determines the sharpening time, which in general can be distinct from the dynamical exponent.

Interesting directions for future work are to explore the application of these new concepts of strong-to-weak symmetry breaking to Hilbert space fragmentation, that fall outside the scope of field-theoretical approaches and are not captured by hydrodynamic descriptions. This could allow for unconventional types of phase transitions in monitored circuits.
Furthermore, we emphasize that, although the phase diagram for the quantum Lifshitz model is well established in 2+1D, the field theory is not able to exclude different scenarios such as first order phase transitions, with condensation of bosons and dipoles at the same point for the same transition value~\cite{Philip}. Understanding the implications of such first-order transitions for the quantum information flow is another exciting future direction.

\textbf{Acknowledgments.--}We thank Sanjay Moudgalya, Leo Radzihovsky, Romain Vasseur,  Philip Zechmann, and Niklas Ziereis for fruitful discussions. We acknowledge support from the Deutsche Forschungsgemeinschaft (DFG, German Research Foundation) under Germany’s Excellence Strategy--EXC--2111--390814868, TRR 360 – 492547816 and DFG grants No. KN1254/1-2, KN1254/2-1, the European Research Council (ERC) under the European Union’s Horizon 2020 research and innovation programme (grant agreement No. 851161 and No. 771537), the European Union (ERC, DynaQuant, No. 101169765), as well as the Munich Quantum Valley, which is supported by the Bavarian state government with funds from the Hightech Agenda Bayern Plus.

\textbf{Data availability.--}All data supporting the findings of this study are contained in this manuscript and its supplemental materials.

\appendix
\vspace{1cm}
\begin{center}
    \textbf{End Matter}
\end{center}

\section{Appendix A: Renyi-2 Correlator for charge and dipole}\label{app_a_em}

We study SW-SSB by computing the spatial decay of correlations that are not invariant under the action of the symmetry groups and act on distinct replicas. This thereby generalizes the detection of conventional SSB. 
We can rewrite the Rényi-2 correlator from Eq.~\eqref{eq:renyi2} as $\mathcal C_2(x-y, t) = \lim_{k\to 0} \sum_{\mathbf m} \mathbb E_u \text{Tr}(\rho_{\mathbf{m}}^{\otimes 2k+1} O^\dagger_xO_y\otimes O^\dagger_yO_x \otimes \mathbb 1\otimes  \ldots \otimes\mathbb 1 \mathcal{W}_2^{\otimes k })$, where we defined $\mathcal W_2$ as the swap operator between two different copies, that determines the boundary conditions, and we have introduced $\mathcal{N} = 2k+1$ replicas. In the replicated model, this coincides with the expectation value $\frac{1}{\mathcal{N}(\mathcal{N}-1)}\sum_{a\neq b}\langle O^{\dagger}_{x,a}O_{y,a}O^{\dagger}_{y,b}O_{x,b}\rangle$ where $a,b$ are different replica indexes. When the Rényi-2 correlator does not decay with distance, the system exhibits SW-SSB and is in a weakly symmetric non-equilibrium phase. When the Rényi-2 correlator decays with distance the system is strongly symmetric. 
$O_{x, a}$ are local operators that are charged under the considered symmetry. Explicitly for our case, the operator $O^c_{x,a}\sim e^{i\phi(x,a)}$ creates a charge at point $x$ and replica $a$, while $O^d_{x,a}\sim e^{i\partial_x\phi(x,a)}= e^{i\phi_d(x,a)}$ creates a dipole at $x$ and replica $a$.  

In 1+1 D charge vortices are unbound for any non-zero value of the measurement and the correlator decays exponentially \cite{Zechmann2023, Lake2023, Boesl_2024}. For the dipole, the resulting correlator decays algebraically for small measurement rate (quasi-long range order), and exponentially when  dipole vortices unbind. 
Concretely, we consider the Lagrangian density for the $  \mathbb P$ component in the replica indeces, as $\phi_a-\phi_b$ has zero overlap with the $\mathbb Q$ space. For 1+1 D, we find~\cite{SM} that the effective Lagrangian is given by
$\mathcal L= (\partial _\tau   \mathbb P \varphi)^2 + (\partial _x^2   \mathbb P \varphi)^2- \lambda_1 \cos( \partial _x   \mathbb P \varphi)- m_d \cos (  \mathbb P \varphi)\sim (\partial _\tau   \mathbb P \varphi)^2 + (\partial _x^2   \mathbb P \varphi)^2+ \lambda_1(\partial _x   \mathbb P \varphi)^2- m_d \cos (  \mathbb P \varphi)$; we used here $\lambda_1=g_b$, always relevant in a RG sense, and $m_d =g_s =0$ in the dipole-weakly symmetric phase, $m_d =g_s >0$ in the dipole-strongly symmetric regime. 
A simple calculation introducing sources of the fields gives for the logarithm of the correlator for charge and dipole
\begin{align}
  &  \ln(\mathcal C_2^c(x,t))=\ln (\langle e^{-i\phi_a(x,t)+i\phi_b(x,t)} e^{i\phi_a(0,0)-i\phi_b(0,0)}\rangle)\nonumber\\&\propto-\int dk  dw \frac{w^2(1-\cos(kx-wt))}{(w^2+k^2)^2(w^2+\lambda_1 k^2+m_d)}\\&
    \ln(\mathcal C_2^d(x,t))=  \ln (\langle e^{-i\phi_{d a}(x,t)+i\phi_{db}(x,t)} e^{i\phi_{d,a}(0,0)-i\phi_{d b}(0,0)}\rangle)\nonumber\\&\propto- \int dk  dw \frac{w^2 k^2 (1-\cos(kx-wt))}{(w^2+k^2)^2(w^2+\lambda_1 k^2+m_d)}.
\end{align} 

We consider spatial correlations equal time, $t=0$; for the dipole-weakly symmetric phase  one finds that the correlator decays algebraically for the dipole, with an exponent that depends on the measurement rate through $\lambda_1=g_b$, and exponentially for the charge. In the gapped phase on the other hand, both the correlators for charge and dipole decay exponentially.

In 2+1 D the Renyi-2 correlator defines three different phases \cite{Lake2022}. At low measurement rate, dipole condense, while charge has QLRO; the Renyi-2 correlator then is constant for the dipole, but decays algebraically for the charge. At intermediate measurement rate, only the dipole is condensed and the correlator decays exponentially with distance for the charge. Further increasing the measurement rate determines a transition to a phase where both charge and dipole are non condensed and the Rényi-2 decays exponentially for both of them.    

\section{Appendix B: Density fluctuations for charge and dipole}\label{app_b_em}
As discussed in the main text, the sharpening times of charge and dipole are determined by the gap of the transfer matrix. In our system, charge and dipole are not independent from each other as their densities are related by a derivative. In this section, we compute the time dependence of charge and dipole fluctuations in finite region of the system; a quantity not directly related to the sharpening time and that exhibits distinct scaling compared to the transfer matrix. 

This should be contrasted with quantum circuits with charge conservation only, for which fluctuations decay on typical times that scale linearly with system size in a weakly symmetric regime, and logarithmically in the strongly symmetric regime. Hence, the gap of the transfer matrix and density fluctuations are related in such settings. In our case, however, the unconventional properties of the dipole condensed phase, which is a \textit{compressible} gapped state for the charge degrees of freedom~\cite{Lake2023,Zechmann2023}, allow for slower suppression of density fluctuations than the dephasing time of a charged operator that is determined from the transfer matrix gap and sets the sharpening time scales.

In the replica formalism, the spatial fluctuations on a subregion $\ell$ of the system, conditioned on measurement outcomes of charge or dipole density can be written as $\mathbb \sum_{x, y \in \ell} E_{\mathbf m}[\langle n^{c,d} (x, t)n^{c,d} (y, t) \rangle-\langle n^{c,d} (x, t)\rangle\langle n^{c,d} (y,t)\rangle ]= \sum_{x, y \in \ell} \lim _{\mathcal{N}\to 1}\frac{1}{\mathcal{N}(\mathcal{N}-1)}   \mathbb P _{ab}\langle n^{c,d} _a(x,t) n^{c,d}_b(y,t)\rangle $ \cite{Agrawal2022, Barrat2022}. Using hydrodynamic assumptions, we express the evolution of density excitations as a kernel,
$
      \mathbb P _{a, b }\langle n^{c,d} _a(x, t) n^{c,d}_b(0, 0)\rangle= \int dx_1 K (x_1-x, t)   \mathbb P _{a, b }\langle n^{c,d} _a(x_1, 0) n^{c,d}_b(0, 0)\rangle\sim K(x, t)
$
where in the initial state short range correlations of the density fluctuations are assumed. For long  evolution time $t$, this quantity can be connected to the spatial correlations
$
      \mathbb P _{a, b }\langle n^{c,d} _a(x, t) n^{c,d}_b(y,t)\rangle= \int dx_1dy_1 K (x_1-x, t) K (y_1-y )  \mathbb P _{a, b }\langle n^{c,d} _a(x_1, 0) n^{c,d}_b(y_1, 0)\rangle\sim K(x-y, 2t)
$
This argument gives an expression for the variance in terms of the evolution term $\sigma^2_{c,d}(\ell, t)=\ell \int dxK(x,2t)$.
In 1+1 D, we 
find $n_c \sim\partial _x^2\varphi$ and dipole density $n_d \sim\partial _x\varphi$. Consistently with the definition above, the densities are related by a derivative.

The dipole-dipole and charge-charge density correlators hence read
\begin{align}
   \mathbb P \langle n_d(r,t )n_d(0,0)\rangle =\int dk \int dw e^{ikr- iwt} \frac {k^2}{w^2+\lambda_1 k^2 +m_d}\\
      \mathbb P \langle n_c(r,t )n_c(0,0)\rangle =\int dk \int dw e^{ikr- iwt} \frac {k^4}{w^2+\lambda_1 k^2 +m_d}
\end{align}
Where $\lambda_1=g_b$ is always relevant, and $m_d=0$ in the weakly symmetric phase for the dipole and relevant in the strongly symmetric phase. 
Using the procedure outlined above, we obtain the asymptotic scaling for the fluctuations in the region $\ell$ at large times. In the dipole-weak charge-strong symmetric phase, we obtain for the dipole, $\sigma^2_d  (\ell , t )\sim  \ell ^2/t^2$ and a typical time of $\tau^d \sim \ell  $ and for the charge $\sigma^2_c  (\ell , t )\sim  \ell ^2/t^4$ and a typical  time of $\tau^c \sim \ell^{1/2}  $. The polynomial scaling of the typical charge time $\tau^c$ in the strongly symmetric regime for the charge may seem surprising at first. However, this is only because dipole and charge densities are not independent, but related by a derivative. 

In the dipole-strongly symmetric regime,  we obtain the expected exponential decay, $\sigma^2_d  (\ell , t )\sim  \ell e^{-\sqrt m_d t}$, with a typical time of $\tau^d \sim \ln\ell  $ and a similar scaling for the charge fluctuations. 

We proceed similarly for 2+1 D. In this case, the behavior of the $  \mathbb P $ component of the fields is described by the Lagrangian  
\begin{align}\label{radlag}
    \mathcal{L} &= 
      \mathbb{P} 
    [ K_{ijkl} \partial_i \phi_d^j\partial_k\phi_d^l ] 
 + \frac{b_p}{2} \mathbb{P} 
    (\partial_\tau \phi_d)^2+ \frac{c_p}{2} \mathbb{P} 
    (\partial_\tau \phi)^2\nonumber \\&
     + r \frac{d_p}{2} \mathbb{P} 
    (\nabla \phi + \phi_d)^2,
\end{align} as reported in the main text. In the weakly symmetric regime for both charge and dipole, the charge and dipole fields are gapless and vortex excitations are not relevant. For this reason, the system is well described by a Lifshitz-type Lagrangian density $\mathcal L_{ww} =  \mathbb P [\frac {c_p}{ 2} (\partial _t \phi)^2+ K_{ijkl } \partial_i \partial_j \phi \partial_k \partial_l \phi] $. This determines fluctuations that decay in typical times that scale polynomially with the linear size of the system $\tau^c \sim \ell^2 $ and  $\tau^d \sim \ell^4 $. 

In the dipole-weakly, charge-strongly symmetric regime, the charge phase field theory can be integrated out from the Lagrangian~\eqref{radlag}, $ L _{s w} =  \mathbb P[(\partial _t \phi_d)^2+ K_{ijkl } \partial_i  \phi_d^j \partial_k  \phi_d^l ]$ \cite{Radzihovsky2022}. Also in this regime fluctuations are expected to decay polynomially with the linear size of the region $\ell$. We find that for large times the asymptotic behavior of the fluctuations is $\sigma^c (\ell, t)\sim \ell^{4}/t^{5} $ and  $\sigma^d (\ell , t ) \sim \ell^{4}/t^{3} $. In the strongly symmetric regime, both the charge and dipole fields are gapped and fluctuations are suppressed in typical times that scale logarithmic with $\ell$.

In summary, the time scales on which density fluctuations in a subsystem decay are not related to the sharpening time scales. Instead, the sharpening times of charge and dipole degrees of freedom are determined by time-dependent correlation function that measure the response to modifying the charge (i.e., they are response functions of charged operators). The sharpening time scales are thereby determined by the respective gaps of the transfer matrix, as emphasized in the main text.

\bibliography{bib}

\section{  Transfer matrix for the stat mech model  }\label{Loc}
In this section, we obtain an  effective statistical mechanics model for the charge dynamics in the quantum channel, following~\cite{Barratt2022}. As depicted in Fig.~\ref{fig:2} of the main text, the initial density matrix evolves through a strongly symmetric quantum channel $\rho(\mathbf m )=T_{\mathbf m}\rho_0T_{\mathbf m }/p(\mathbf m )$, where $\mathbf m$ is the set of recorded measurements of the charge in space-time and $p(\mathbf m )$ is the associated Born probability. The operator $T_\mathbf m$ is composed of a unitary part, a measurement term and a noise term:
$
T_{\mathbf{m}} = \prod_{t=t_i}^{t_f} T(\mathbf{m}_t), \quad T(\mathbf{m}_t) \cdot T^\dagger(\mathbf{m}_t)=  \mathcal{E}[U_t\mathcal{M}(\mathbf{m}_t)\cdot \mathcal{M}^\dagger(\mathbf{m}_t)U_t^\dagger]
$.
Here, $U_t$ is a unitary gate randomly drawn from the Haar ensemble, $\mathcal{M}(\mathbf{m}_t)$ applies the measurement record at time $t$, and $\mathcal E$ is a noise channel; see  Fig.~\ref{fig:2}. We consider a completely dephasing channel, $\mathcal E[\cdot]=\sum_i K_i^\dagger\cdot K_i $, where $K_i=\ketbra{i}{i}$ is a projector in the local charge basis.

The unitary part of the evolution is a product of gates acting on $n$ neighboring sites $U_t=\otimes_{\mathbf{r}} U_{\mathbf{r}}$, where $\mathbf{r}$ specifies the location of the unitary. A gate acting on $n$ neighboring sites starting at site $i$ takes the form $U_{\mathbf{r}} = \bigoplus_{s \in \mathcal{S}} U^{(s)} P_s$, where $\mathcal{S}$ is the set of connected charge-dipole sectors, $P_s$ is the projector onto sector $s$, and $U^{(s)}$ is a Haar-random unitary of dimension $d_s$. Here, $d_s$ is the size of the charge and dipole sector $s$ and counts the number of connected configurations. This construction naturally generalizes to systems with higher-moment (multipole) symmetries~\cite{Feldermaier2020} or other sets of constrained configurations.

After each layer of unitaries, projective measurements are applied independently yielding a set of recorded measurements in space time $\mathbf m$ that collapse the local charge degree of freedom. Measurements are performed with a rate $\gamma$, that will serve as a control parameter to drive the transition from weak to strong  symmetric non-equilibrium phases.
To compute averaged observables acting on non-linear functions of the density matrix, we derive a statistical mechanics description of the circuit $T_\mathbf{m}$ on replicas. We remark that the same effective model can be obtained also by introducing a monitored circuit of qubit plus qudit degrees of freedom on each site, that evolves conserving total charge and dipole only for the qubits, see following section.
Using a duplicated Hilbert space formalism ~\cite{Barratt2022}, we  can write the averaged channel after a single time step $\delta t$ as $ \mathbb E_u[(T_{\mathbf m}(\delta t)\rho T^\dagger_{\mathbf m}(\delta t))]=\mathbb E_u[\sum_{k} |k\rangle\rangle \langle\langle k| U_{\delta t}\mathcal{M}_\mathbf m \otimes U_{\delta t}^*\mathcal{M}^*_\mathbf m|\rho\rangle \rangle]$. The averaged channel $\mathbb E_u[T_{\mathbf m} \otimes  T^\dagger_{\mathbf m}]$ is hence the relevant quantity. At each space-time position we then average the unitaries $U(x, t)=\bigoplus_{s \in \mathcal{S}} U^{(s)}(x,t) P_s$ , and obtain $\mathbb E_u [U(x, t)\otimes U(x,t)^*]=\sum_s W_s P_s\otimes P_s$. In our model we consider that the circuit either connects one configuration or two, which simplifies the analysis without modifying the universal properties of the circuit. Thus, we obtain $W_s=1$ if $d_s=1$ and $W_s=1/2 \begin{pmatrix}1&1\\1&1\end{pmatrix}\otimes \begin{pmatrix}1&1\\1&1\end{pmatrix}$ for $d_s=2$. The dephasing channel selects the diagonal degrees of freedom, $|k\rangle \rangle=\ket{k}\otimes\ket{k}$, and the averaged operator for the evolution can be replaced by $T_u=\sum_\alpha P_\alpha( G_\alpha+G_\alpha^\dagger)P_\alpha$, where$ ( G_\alpha+G_\alpha^\dagger)=1$ for $d_s=1$,$ ( G_\alpha+G_\alpha^\dagger)=\begin{pmatrix} 1/2& 1/2\\ 1/2& 1/2\end{pmatrix}$ for $d_s=2$ . 

\section{Equivalent set up with infinite local Hilbert space dimension qudits}
The effective transfer matrix evolution described in the main text can be obtained also in a different setting when considering  a monitored quantum circuit defined on a set of qubits and qudits~\cite{Agrawal2022,Barrat2022}. Similarly to the model considered in the main text, the circuit is composed of dipole conserving gates drawn from the Haar ensemble, $
U_{\mathbf{r}} = \bigoplus_{s \in \mathcal{S}} U^{(s)} P_s$,
where $\mathcal{S}$ is the set of connected charge-dipole sectors, $P_s$ is the projector onto sector $s$, and $U^{(s)}$ is a Haar-random unitary of dimension $N = d_s \cdot q^n$. Here, $d_s$ is the dimension of sector $s$ for the qubit, and $q$ is the on-site Hilbert space dimension of the qudit. 
After each brickwork layer of unitary gates, we perform projective measurements independently on each site with rate $\gamma$. These measurements collapse the charge degree of freedom locally on both the qubit and qudit. 
Contrarily to the model considered in the main text, we do not need an additional completely dephasing channel to obtain an analytically tractable model.

As above, we introduce a replicated set of the circuit to study observables that are nonlinear in the density matrix. Each measurement effectively pins the charge values at the corresponding space-time location across all replicas. The non-equilibrium phases are characterized by the R\'enyi-2 correlator and can be computed from the effective statistical mechanics model.
We use standard techniques from Haar integration and Weingarten calculus to derive an effective statistical model for the charge degrees of freedom. To compute averaged $\mathcal{N}$-th moments of the density matrix, we introduce $\mathcal{N}$ forward and $\mathcal{N}$ backward replicas of the system, and then average over the randomness in both the unitary circuit and the measurement outcomes. In particular, averaging over Haar-distributed unitaries acting within each symmetry sector replaces the replicated unitary product $U_1 U_2 \dots U_{\mathcal{N}} U_{\mathcal{N}}^* \dots U_2^* U_1^*$ with a weighted sum over products of permutations in the replicas, $\sigma_{\mathcal{N}}, \sigma'_{\mathcal{N}} \in \mathcal{S}_{\mathcal{N}}$, the symmetric group on $\mathcal{N}$ elements.
This procedure yields a statistical mechanics model where each site hosts a pair of permutation-valued spins, representing elements of the symmetric group between replicas. The model assigns weights to each configuration based on the structure of the connected symmetry sectors. Importantly, the conserved quantities influence the size of each connected sector, thereby entering the model via the sector-dependent weights.

To simplify the resulting model, we consider the limit of infinite local Hilbert space dimension, $q \to \infty$. In this limit, configurations with differing permutations on the same site are suppressed, effectively locking the two permutations into a single degree of freedom.

On each leg of the circuit, a charge degree of freedom (associated with the qubit) evolves according to dipole-conserving dynamics on the square lattice. As previously discussed, the charge configuration entering a vertex determines the local vertex weight, given by $D_a = 1 / (d_s q^5)$. Hence, all possible charge configurations in the effective statistical model can be described by an imaginary-time evolution generated by a transfer matrix. The specific form of this matrix depends on the structure of the connected configuration sectors. 

Upon integrating out the permutation degrees of freedom, we obtain a transfer matrix for the charge dynamics encoded in the qubits, that connects all the states belonging to the same symmetry sector with equal probability $1/d_s$. 

\section{Field Theory Model for 1+1D}
In this section we explicitly derive the field theory to describe phase transitions of the 1+1D model. In order to formally avoid Hilbert space fragmentation in the circuit we consider a set of dipole conserving gates that act on five sites. For simplicity, we consider the set of gates $G = $
\begin{tikzpicture}
  \draw[pink, very thick] (0,0) rectangle (2.5,0.5);
  \draw [black ] (0.25,0.25) circle (0.15);
  \draw [black, fill=black ] (0.75,0.25) circle (0.15);
   \draw [black, fill=gray ] (1.25,0.25) circle (0.15);
  \draw [black , fill=black] (1.75,0.25) circle (0.15);
  \draw [black ] (2.25,0.25) circle (0.15);
  \draw[ultra thick, ->] (0.75,0.5) arc (0:150:0.3);
    \draw[ultra thick, ->] (1.75,0.5) arc (180:30:0.3);
\end{tikzpicture}
and $G^\dagger=$ \begin{tikzpicture}
  \draw[pink, very thick] (0,0) rectangle (2.5,0.5);
  \draw [black , fill=black] (0.25,0.25) circle (0.15);
  \draw [black] (0.75,0.25) circle (0.15);
   \draw [black, fill=gray ] (1.25,0.25) circle (0.15);  
  \draw [black ] (1.75,0.25) circle (0.15);
  \draw [black , fill=black] (2.25,0.25) circle (0.15);
  \draw[ultra thick, ->] (0.25,0.5) arc (180:30:0.3);
    \draw[ultra thick, ->] (2.25,0.5) arc (0:150:0.3);
\end{tikzpicture}
where the gray circle indicates a configuration that can be either full or empty, black is filled, and white is empty. We define the projector $P$ over the configurations where a dipole-hopping can be applied.
We write the  transfer matrix of the model considering a probability $J$ to apply a dipole-hopping gate and a  unit time d$t$  when the transfer matrix is applied,  
\begin{align}
    T&=\mathbb{1}(\mathbb 1 -P)+ \{ J\, dt (G+G^\dagger)+ (1-Jdt)\mathbb 1\} P\nonumber\\
    &=\mathbb 1 + J\, dt  (G+G^\dagger)P-P\, dt\nonumber \\
    &\sim \text{exp}(dt J((G+G^\dagger)P-P))
\end{align}
We have obtained an effective Hamiltonian for the evolution of the model
$H=-J((G+G^\dagger )P-P)$. Furthermore, we consider smooth measurements on each qubits that enforce the same outcome of the measurements on every replicas
\begin{equation}
    H_m(\tau)=\frac \gamma 2 \sum_{i,a} (\sigma_{i,a}^z -m_i(\tau))^2=\frac \gamma 2 \sum_{i,a, b } \sigma_{i,a}^z \mathbb P_{a,b} \sigma_{i,b}^z
\end{equation}
where $i$ gives the lattice site and $a$ the replica index. The field $m_i$ coincides with the ``softened" measurements  outcome, $m_i\in [-1,1]$. The second equality is obtained by integrating out the measurements outcomes and $\mathbb P_{a,b}=\delta_{a,b}-1/\mathcal N$ is a projector in the replica degrees of freedom, and $\mathcal N$ is the number of replicas. The Hamiltonian for the total model reads:
\begin{widetext}
\begin{align}
H=&-J\sum_{i,a}\bigg\{ 2 \big [( \bar S_{i,a} \cdot \bar S_{i+1,a}- S_{i,a}^zS_{i+1,a}^z)\otimes ( \bar S_{i+3,a} \cdot \bar S_{i+4,a}- S_{i+3,a}^zS_{i+4,a}^z)+(\bar{S}_{i,a}\times \bar{S}_{i+1,a})|_z \otimes (\bar{S}_{i+3,a}\times \bar{S}_{i+4,a})|_z]\nonumber \\& -\big[\frac 1 8 \ +\frac 1 2 (- S^z_{i,a}S^z_{i+1,a}-S^z_{i,a}S^z_{i+3,a}+ S^z_{i,a}S^z_{i+4,a}+S^z_{i+1,a}S^z_{i+3,a}-S^z_{i+1,a}S^z_{i+4,a}-S^z_{i+3,a}S^z_{i+4,a}+\nonumber \\&+2S^z_{i}S^z_{i+1}S^z_{i+3} S^z_{i+4}\big] \bigg\}+ 2\gamma \sum_{a,b } S^z_a\mathbb P _{a,b} S^z_b .
\end{align} We can now introduce a spin path integral formalism. To do this, we consider set of coherent spin states $\ket{\mb n }=e^{i \theta (\mathbf  n_0\times \mathbf n) \mathbf{S}} $, where the spin operators act as $\bra{\mb n } \mb S \ket{\mb n}= S \mb n $ , $\mb n =(\sin\theta\cos\phi , \sin\theta\sin\phi , \cos\theta)$ and $\mathbf n_0= (0, 0 , 1)$~\cite{Fradkin_2013}. We obtain
\begin{align}\label{Full_Lagrangian}
   \mathbb L=&-\frac{i}{2} \sum_{i,a} \cos\theta_{a,i}\partial_\tau \phi_{a,i} - J\sum_{i,a}\bigg\{ \frac{1}{8} \big[\sin{\theta_{i,a}}\sin{\theta_{i+1,a}}\sin{\theta_{i+3,a}}\sin{\theta_{i+4,a}} \cos (\phi_{i,a}-\phi_{i+1,a}-\phi_{i+3,a}+\phi_{i+4,a})\big] \nonumber \\&
    -\big[\frac 1 8- \frac 1 8 \cos{\theta_{i,a}}\cos{\theta_{i+1,a}}- \frac 1 8 \cos{\theta_{i,a}}\cos{\theta_{i+3,a}+ \frac 1 8 \cos{\theta_{i,a}}\cos{\theta_{i+4,a}}+ \frac 1 8 \cos{\theta_{i+1,a}}\cos{\theta_{i+3,a}}}- \frac 1 8 \cos{\theta_{i+1,a}}\cos{\theta_{i+4,a}}+\nonumber \\&- \frac 1 8 \cos{\theta_{i+3,a}}\cos{\theta_{i+4,a}}+ \frac 1 8 \cos{\theta_{i,a}}\cos{\theta_{i+1,a}}\cos{\theta_{i+3,a}}\cos{\theta_{i+4,a}}\big]\bigg\}+ \frac{\gamma}{2} \sum_{i, a,b} \cos{\theta_{i,a}} \mathbb P_{a,b}\cos{\theta_{i,b}}.
\end{align} \end{widetext} 
The field theory obtained with the functional field integral should be consistent with bosonization techniques. To proceed, we notice that several non-equilibrium phases can be induced by the measurements. Generally this happens in two stages. First, sectors with very different quantum numbers can be easily distinguished by the monitored dynamics, but nearby sectors, say $N$ and $N+1$, remain unresolved. Second, only at a later stage those nearby charges will resolved. From those sectors the one at half-filling (i.e., zero magnetization) is the larges and dominates the process. It is then justified to consider the charge as a  perturbation around half filling. This is equivalent to performing an expansion of the angle $\theta=\frac \pi 2 +\vartheta$.  Replacing the discrete lattice sites with vectors and taking the lattice derivative, we obtain a Lagrangian density 
\begin{align}
    \mathcal L=&\sum_a  \frac i 2 \vartheta_a(x)\partial_{\tau} \phi_a (x) +\frac {9J}{16} \big\{ (\partial_x ^2 \phi_a(x))^2+ (\partial_x ^2 \vartheta_a(x))^2\}+\nonumber \\&+\frac \gamma 2 \sum_{a,b} \vartheta_a \mathbb P_{a,b} \vartheta _b,
\end{align}
which we obtained by expanding the cosine and sine terms. Here, we momentarily discard the effect of vortices; they will be added later, while studying the possible instabilities driven by excitations. The effect of vortices can also be analyzed starting from the full Lagrangian \ref{Full_Lagrangian} and using a Villainization procedure as done in \cite{Lake2023}; the two approaches give the same result. We integrate over $\bar \vartheta_a= \mathbb Q_{a,b} \vartheta_b$. Taking a long wave-length approximation and rescaling space and time variables to obtain an isotropic action,  we find the Lagrangian density
\begin{align}\label{eq:6}
    \mathcal L=&\sum_a\frac{i}{2} \bar \vartheta_a \partial_\tau \bar \phi_a + \frac {\bar \rho }{2} \big[(\partial_x^2 \bar \vartheta_a)^2+(\partial_x^2 \bar \phi)^2))\big] +\nonumber\\&\frac{\rho_s}{2} \{(\partial^2_x (\mathbb P \phi)_a)^2+(\partial_\tau(\mathbb P\phi)_a )^2\},
\end{align}
where we introduced the parameters $\bar \rho =\frac{9J}{16}$ and $\rho_s=1/4(\frac{9J}{16\gamma^2})^{1/3}$.
We observe, that this procedure has lead us to the Lifschitz model for the phase field $\mathbb P \phi$. In 1+1D dimension this model does not undergo any phase transition for the charge degrees of freedom. Then, we expect that for any measurements rate, the system will collapse in a definite charge sector after times that scale logarithmically in system size. However this does not give us any information about the dipole sector. 
To obtain further insights on the behavior of the dipole, we introduce an effective field theory for the dipole degrees of freedom, using the relations between density and phase of charge and dipoles obtained via bosonization procedures, $\theta =\partial_x \theta_d$ and $-\partial_x \phi=\phi_d$~\cite{Lake2022,Zechmann2023}. We write the effective field theory 
\begin{align}
    \mathcal{L}=&\frac{i}{2}\bar \vartheta\partial_\tau\bar \phi + \frac{\bar \rho }{2} (\partial_x^2 \bar \vartheta)^2 + \big[\big(\frac{\rho_s}{2} \mathbb P+ \frac{\bar \rho}{2}\mathbb Q \big)\partial_x \phi_d]\partial_x \phi_d +\nonumber \\& +  \big[\big(\frac{\rho_s}{2} \mathbb P+ \frac{\bar \rho}{2}\mathbb Q \big)\partial_\tau \phi_d]\partial_\tau \phi_d + r\big[\big(\frac{\rho_s}{2} \mathbb P+ \nonumber \\&+\frac{\bar \rho}{2}\mathbb Q \big)(\partial_x \phi +\phi_d)](\partial_x \phi +\phi_d),
\end{align}
where $\mathbb P \phi \phi$ has to be interpreted as a scalar product $\mathbb P \phi \phi=\sum_{a,b } \mathbb P_{a,b} \phi_{b}\phi_{a}=\mathbb P \phi \mathbb P\phi $. As explained in the main text, in this effective field theory the dipole $\phi_d$ and charge $\phi$ degrees of freedom are formally independent. We enforce the usual equality for the phase field  $\phi_d=-\partial_x\phi $  via a Higgsing procedure, since for large values of the parameter $r$ all the field configurations that do not satisfy the relation $\phi_d=-\partial_x\phi $ are strongly gapped. The original Lagrangian $\mathcal L$ can be obtained by integrating out the dipole field $\phi_d$ and taking the large $r$ limit. Similarly an effective Lagrangian for the dipole field $\mathcal L_d$ can be obtained by integrating out $\phi$ and taking the large $r$ limit.

\section{Introducing vortex excitations}
After having obtained the model, we want to study its phase diagram and in particular the instabilities driven by vortex proliferation. To this end, we decompose the fields in a smooth and a vortex part both for charge $\phi=\phi^s+\phi^v$ and dipole $\phi_d=\phi_d^s+\phi^v_d$. In what follows, we will use a similar strategy as the ones that is considered in 2D classical smectic theories, see for example~\cite{Pretko2019,Zhai2019,Radzihovsky2020,Zhai2021}. In order to decouple  these components, we perform a Hubbard-Stratonovich transformation, introducing two new fields $\sigma=(\sigma_\tau, \sigma_x)$ and  $j=(j_\tau, j_x)$. Furthermore, we define $\sigma_\tau=\bar \vartheta/2$. We obtain
\begin{align}
    \mathcal L &=i \sigma_\tau\partial_\tau(\phi^v+\phi^s) + i\sigma_x(\partial_x \phi^s+\phi^s_d+\partial_x \phi^v+\phi^v_d)+\nonumber\\&i (j_\tau \partial_\tau + j_x\partial_x)(\phi^s_d+ \phi^v_d )+\frac{\sigma_\tau^2}{2\rho_s}+ \frac{1}{2 r} (\frac{1}{\rho_s}\mathbb P + \frac{1}{\bar \rho} \mathbb Q)\sigma_x \sigma_x+ \nonumber\\&+\frac{1}{2 } (\frac{1}{\rho_s}\mathbb P + \frac{1}{\bar \rho} \mathbb Q)j_\tau j_\tau +\frac{1}{2 } (\frac{1}{\rho_s}\mathbb P + \frac{1}{\bar \rho} \mathbb Q)j_x j_x .
\end{align}
Integrating out the smooth component of the fields $\phi^s$ and $\phi_d^s$ we obtain a system of coupled differential equations 
\begin{equation}
\begin{cases}
        \partial_\tau \sigma_\tau+\partial_x\sigma_x=0\\       
        \partial_\tau j_\tau +\partial_x j_x-\sigma_x=0
\end{cases}
\end{equation}
We solve first for $\sigma$, introducing the new field $\sigma_i =\epsilon_{i,j} \partial_j \alpha / 2\pi $. The second equation is then solved when we consider $j_i=\epsilon_{i,j} \partial_j \varphi /2 \pi +\epsilon_{x,i} \alpha/2\pi$, which implies $j_\tau=\partial_x \varphi/2\pi-\alpha/2\pi$ and $j_x=-\partial_\tau \varphi /2 \pi$. Substituting these solutions, we obtain
\begin{align}
    \mathcal L =&i \frac{\alpha}{2\pi} b + i \frac{\varphi}{2\pi} s + \frac{1}{8\pi^2 \rho_s} (\partial_x \alpha)^2+ 2\bar{\rho} (\partial_x^3\alpha)^2+\nonumber\\&+\frac{1}{8\pi^2  }\bigg\{(\frac{1}{\rho_s} \mathbb P + \frac{1}{\bar \rho} \mathbb Q)(\partial_x \varphi -\alpha )(\partial_x \varphi -\alpha )+\nonumber\\&+(\frac{1}{\rho_s} \mathbb P + \frac{1}{\bar \rho} \mathbb Q)(\partial_\tau \varphi  )(\partial_\tau \varphi  )\bigg\}+\frac{1}{8\pi^2 r }(\frac{1}{\rho_s} \mathbb P + \nonumber\\&\frac{1}{\bar \rho} \mathbb Q)(\partial_\tau \alpha )(\partial_\tau \alpha ),
\end{align}
where we defined the vortex densities for the charge and dipole respectively.
\begin{align}
    b=\epsilon_{i,j}\partial_i \partial_j \phi^v=2\pi\sum_i b_i\delta (\bold r-\bold r_i),\\
    s=\epsilon_{i,j}\partial_i \partial_j \phi^v_d=2\pi\sum_i s_i\delta (\bold r-\bold r_i).
\end{align}
We highlight that, in this picture, the field $\varphi$ is conjugate with the dipole vortices $s$, and as such $e^{i\varphi}$ inserts a vortex for the dipole. On the other hand, $\alpha$ is conjugate to the charge vortex density. Hence $e^{i\alpha}$ creates the charge vortices. Here, $\varphi $ and $\alpha$ are related by a Higgsing procedure, as taking the large $r$ limit gaps all the field configurations that do not satisfy $\partial_x \varphi=\alpha$.
Integrating out $\alpha$ and taking the large $r$ limit, we find a Lagrangian for $\varphi$
\begin{align}
    \mathcal L&= \frac{1}{8 \pi^2 \rho_s}\bigg[(\partial_x^2 \mathbb P\varphi)^2+(\partial_\tau \mathbb P\varphi)^2\bigg]+i\frac{\varphi }{2 \pi}s + i \frac{\partial_x \varphi}{2 \pi} b + \nonumber\\& +\frac{\rho_s}{2 }\mathbb P b b + \frac{1}{8 \pi^2 \bar \rho } \big[(\mathbb Q \partial_\tau \varphi)^2+2\bar \rho ^2 (\partial ^4_x \mathbb Q \varphi )^2 \big]+\frac{\bar \rho}{2}\mathbb Q b b.
\end{align}
To complete the description of the vortex excitations, we add by hand the core energy of the charge and dipole degrees of freedom,
\begin{align}
    \mathcal L&= \frac{1}{8 \pi^2 \rho_s}\bigg[(\partial_x^2 \mathbb P\varphi)^2+(\partial_\tau \mathbb P\varphi)^2\bigg]+i\frac{\varphi }{2 \pi}s + i \frac{\partial_x \varphi}{2 \pi} b + \nonumber\\& +\frac{\rho_s}{2 }\mathbb P b b + \frac{1}{8 \pi^2 \bar \rho } \big[(\mathbb Q \partial_\tau \varphi)^2+2\bar \rho ^2 (\partial ^4_x \mathbb Q \varphi )^2 \big]+ \nonumber\\& +\frac{\bar \rho}{2}\mathbb Q b b+ E_b b^2 + E_s s^2.
\end{align}
From now on we focus on the fields that contribute to the non-linear averages and can potentially drive a non-equilibrium transition for the dipole. Then, we want to study the possible transitions driven in the $\mathbb P$ components of the field, 
\begin{align}
    \mathcal L&= \frac{1}{8 \pi^2 \rho_s}\bigg[(\partial_x^2 \mathbb P\varphi)^2+(\partial_\tau \mathbb P\varphi)^2\bigg]+i\frac{\varphi }{2 \pi}s + i \frac{\partial_x \varphi}{2 \pi} b + \nonumber\\& +(\frac{\rho_s}{2 }+E_b) b b +   E_s s s.
\end{align}
Furthermore, the field $\bar \phi$ does not undergo a non-equilibrium transition, neither of the charge nor of the dipole; this is discussed in depth and proven in the next-next section. These considerations allows us to focus on vortex configurations that have zero components in the $\mathbb Q$ space and can thus unbind.

\section{Mapping to sine-Gordon model}
Following the steps of \cite{Zhai2021}, in order to proceed we want to write a sine-Gordon formulation of this model. We define 
\begin{align}
    S_0&=\int \text{dx} \, \text{d}\tau \, \frac{1}{8 \pi^2 \rho_s}\bigg[(\partial_x^2 \mathbb P\varphi)^2+(\partial_\tau \mathbb P\varphi)^2\bigg],
    \\S_1&=\int \text{dx} \, \text{d}\tau \,i\frac{\mathbb P\varphi }{2 \pi}s + i \frac{\partial_x \mathbb P \varphi}{2 \pi} b   +(\frac{\rho_s}{2 }+E_b)\mathbb P b b  + E_s \mathbb Ps s=\nonumber\\&=\sum_{i_s, i_b} \,i\frac{\mathbb P\varphi (\bold r_{i_s}) }{2 \pi}s_{i_s} + i \frac{\partial_x \mathbb P \varphi(\bold r_{i_b})}{2 \pi} b_{i_b} +\nonumber \\& +(\frac{\rho_s}{2 }+E_b)\mathbb P b_{i_p} b_{i_p}  + E_s \mathbb Ps_{i_s} s_{i_s},
\end{align} where the second equality for $S_1$ follows after performing explicitly the integral over the vortex density. The partition function of the model is then 
\begin{align}
  \mathcal Z=\int D[\mathbb P \varphi]  e^{-S_0[\mathbb P \varphi]}\sum_{c \in \mathcal C} e^{ S_1^c[ \mathbb P \varphi]},
\end{align}
where we defined $\mathcal{C}$ as the set of all possible configurations of the vortices. 
The minimal charge vortex configurations in $\mathbb P $, is obtained for vortex configurations where there exists two replicas $a\not= b $ such that $( b_{i_b})_a= 1$ and $( b_{i_b})_b=-1$, and same for the dipole vortices $s$. Indeed, these configurations have zero total vorticity, and as such zero overlap with $\mathbb Q$. 
\begin{widetext}
\begin{align*}
  \mathcal Z=\int D[\mathbb P \varphi]  e^{-S_0[\mathbb P \varphi]}\bigg[ 1+ \sum_{a \not= b} \int d \mb r e^{-2E_b}e^{i( \partial_x\varphi_a -\partial_x\varphi_b)}+ \sum_{a \not= b< c \not= d} \int d \mb r_1 \int d \mb r_2 e^{-4E_b}e^{i( \partial_x\varphi_a -\partial_x\varphi_b+\partial_x \varphi_c -\partial_x \varphi_d)}\ldots\bigg]\\\bigg[ 1+ \sum_{a \not= b} e^{-2E_s}e^{i(  \varphi_a - \varphi_b)}+ \sum_{a \not= b< c \not= d} \int d \mb r_1 \int d \mb r_2 e^{-4E_s}e^{i( \varphi_a -\varphi_b+\varphi_c -\varphi_d)}+\ldots\bigg]\\
\end{align*}
\end{widetext}
We then obtain the effective Lagrangian
\begin{align*}
    &\mathcal L=\frac{1}{8 \pi^2 \rho_s}\bigg[(\partial_x^2 \mathbb P\varphi)^2+(\partial_\tau \mathbb P\varphi)^2\bigg]-\\& \sum_{a\not= b \,\, m} [ g_b^{(m)} \cos(m(\partial_x \varphi_a-\partial_x \varphi_b))+g_s^{(m)} \cos(m( \varphi_a- \varphi_b))]
\end{align*}
with $g_b^{(m)}=e^{-2m(E_b+\rho_s/2)}$ and  $g_s^{(m)}=e^{-2mE_s}$. Since the prefactors $g_{b,s^{(m)}}$ decay exponentially with  $ m$, in the following we consider only its lowest values. This result in the replicated theory is equivalent to the one obtained for a single copy by a different method of Villainization in \cite{Lake2023}. We study the relevance of the cosine operators by analyzing the the correlations at the Gaussian fixed point. In particular one finds that $\langle e^{i(\varphi_a (\mb r)-\varphi_b (\mb r))} e^{-i(\varphi_a (\mb 0)-\varphi_b (\mb 0))}\rangle $ decays exponentially. Hence, the operator that creates dipole vortices is irrelevant at the Gaussian point $g_b=0$, $g_s=0$. By contrast, the charge vortex operator $ \langle e^{i(\partial_x\varphi_a (\mb r)-\partial_x \varphi_b (\mb r))} e^{i(\partial_x\varphi_a (\mb 0)-\partial_x\varphi_b (\mb 0))}\rangle  \sim \text{const}$  is always relevant \cite{Zhai2021}. As $\partial_x \varphi_a-\partial_x \varphi_b$ will localize in one of the minima of the cosine, at sufficiently long scales we can thus expand the cosine in a harmonic potential 
\begin{align*}
  \mathcal L &=  \sum_a \bigg[\frac{1}{8 \pi^2 \rho_s}(\partial_\tau \mathbb P\varphi_a)^2-\sum_{b\not= a} \frac{ g_b }{2} (\partial_x \varphi_a (\mb r)-\partial_x \varphi_b (\mb r))^2\\&-g_s \cos( \varphi_a (\mb r)-\varphi_b (\mb r)),
\end{align*}
where we discarded higher-order derivatives in the spatial component. We notice that the rate of measurements still controls the spatial component through the parameter $g_b=e^{-2 (\rho_s+E_b)}$. We can then analyze the transition by computing at the correlation decay given that the screening of the charge-phase vortices has been introduced. We find
\begin{equation}
    \langle e^{i(\varphi_a (\mb r)-\varphi_b (\mb r))} e^{i(\varphi_a (\mb 0)-\varphi_b (\mb 0))}\rangle  \sim 1/r^{2K}
    \end{equation}
and $K= K(\gamma)=\sqrt{2 \pi^2 \rho_s/g_b}=\frac{\pi^2}{2} (\frac{9J}{16\gamma^2})^{1/6}e^{(9J/16\gamma^2)^{1/3}/8+E_b}$. We now use known results from the sine-Gordon theory. The critical value for the transition is given by $K=2$, that determines the critical value of measurements $\gamma_c$; we further notice that $K(\gamma)$ is a monotonously decreasing function,   $K \to +\infty $ for $\gamma\to 0$ and $K \to 0 $ for $\gamma\to +\infty$. The value of $\gamma_c$ depends on the core energy of the charge vortices, that cannot be determined from our theory and depends on the specific characteristics of the model considered.

\section{Choice of vortex configurations}
We now show that the minimal vortex configuration that can drive the phase transition is a bounded vortex-antivortex pair acting on different replicas and the same point in space. At the same time, pairs that coexist in the same replica cannot unbind. In particular, these latter configurations of vortices have a non-zero component in the space $\mathbb Q$ since their inter-replica average is non-zero. We compute the decay of correlations for the fields $\bar \phi=\mathbb Q \phi$ to show that the vortices of the phase of the charge do not unbind. In the Lagrangian Eq.~\eqref{eq:6} we integrate out the field $\bar \vartheta$ to obtain a theory only for $\bar \phi$. After Fourier transform, we obtain $\bar{\mathcal L}=\frac{1}{2} \frac{1}{4 \bar \rho}[\omega^2/k^4 + Dk^4]\bar \phi (\mb q) \bar \phi (\mb -q)$, where we introduced the parameter $D=2 \bar \rho^2$. We compute the spatial correlations of the order parameter for the charge $\langle\bar \psi (\mb r) \bar \psi (\mb 0)\rangle=\langle e^{-i q_0(\bar \phi (\mb r)- \bar \phi (\mb 0))}\rangle=e^{-\frac 1 2 q_0^2\langle(\bar \phi (\mb r)- \bar \phi (\mb 0))^2\rangle} \sim \text{const}$, where we use the expectation value 
\begin{align*}
  &  \langle(\bar \phi(\mb s)-\bar \phi(\mb 0))^2\rangle=&\\&=\int \text d \mb q\int \text d \mb s' (e^{i\mb q \mb s}-1)(e^{i\mb q'\mb s}-1)\langle\bar{\phi}(\mb q)\bar{\phi}(\mb q')\rangle \\&
    =\int \text d \mb q (2-2\cos(\mb q \mb s) \frac{4\bar \rho k^4}{w^2+D k^8}  =_{s \to \infty} 2 \sqrt{\pi} \Lambda
\end{align*}
and we defined $\mb q=(w, k)$, $\mb s=(t, \mb r)$, and $\Lambda$ a convenient cutoff. Similarly, such calculations can be done for the order parameter of the dipole. In particular, one finds  $\langle\bar \psi_d (\mb s) \bar \psi_d (\mb 0)\rangle=\langle e^{-i q_0(\bar \phi_d (\mb s)- \bar \phi_d (\mb 0))}\rangle= e^{-\frac 1 2 q_0^2\langle(\bar \phi_d (\mb s)- \bar \phi_d (\mb 0))^2\rangle} \sim e^{-2 \sqrt{\pi} \Lambda^3}$, where we used the defining relation $\phi_d=-\partial_x\phi$. Vortices for charge and dipoles are therefore bound for the field $\bar \phi$ and it is not possible to drive a transition as one changes the measurement rate. In particular, this implies that single replica vortices, that have a component in the inter-replica averaged field, must be bound, since they would otherwise determine a decay of correlations of the phase $\bar \phi$, inconsistent with the above considerations. As such, single-replica charge and dipole bound vortices cannot drive a phase transition. 

Computing correlations of the order parameter for the $\mathbb P$ components of  the field would instead give an exponential decay for the charge vortices, and a constant dependent on the measurement rate $\sim e^{-(\sqrt{\pi }\Lambda/\rho_s)}$  for the dipoles. These correlations show that a dipole vortex unbinding transition can be driven by increasing the rate of measurements. 

Given these considerations, we focus only on vortices in the field $\mathbb P \phi $. The minimal of such configurations are vortex-antivortex pairs in two different replicas, $b_a=-1,\, b_b =1$ and $b_i=0, \, \forall i \not=a,b$.

\end{document}